\documentclass[APL,twocolumn]{revtex4-1}
\usepackage{natbib}
\usepackage{graphicx}
\usepackage{subfigure}
\usepackage{setspace}
\usepackage{amssymb}

\begin{document}

\title{Correlated photon-pair generation in a periodically poled MgO doped stoichiometric lithium tantalate reverse proton exchanged waveguide}
\author{M. Lobino$^1$}
\email[]{mirko.lobino@bristol.ac.uk}
\author{G. D. Marshall$^{2,3}$}
\author{C. Xiong$^{2,4}$}
\author{A. S. Clark$^{1}$}
\author{D. Bonneau$^1$}
\author{C. M. Natarajan$^5$}
\author{M. G. Tanner$^5$}
\author{R. H. Hadfield$^5$}
\author{S. N. Dorenbos$^6$}
\author{T. Zijlstra$^6$}
\author{V. Zwiller$^6$}
\author{M. Marangoni$^7$}
\author{R. Ramponi$^7$}
\author{M. G. Thompson$^1$}
\author{B. J. Eggleton$^{2,4}$}
\author{J. L. O'Brien$^1$}
\affiliation{$^1$Centre for Quantum Photonics, H. H. Wills Physics Laboratory \& Department of Electrical and Electronic Engineering, University of Bristol, Merchant Venturers Building, Woodland Road, Bristol, BS8 1UB, UK}
\affiliation{$^2$Centre for Ultrahigh bandwidth Devices for Optical Systems (CUDOS), Australia}
\affiliation{$^3$MQ Photonics Research Centre, Dept. of Physics and Astronomy, Macquarie University, NSW 2109, Australia}
\affiliation{$^4$Institute for Photonics and Optical Science (IPOS), School of Physics, University of Sydney, Camperdown NSW 2006, Australia}
\affiliation{$^5$Scottish Universities Physics Alliance and School of Engineering and Physical Sciences, Heriot-Watt University, Edinburgh, EH14 4AS, United Kingdom}
\affiliation{$^6$Kavli Institute of Nanoscience, TU Delft, 2628CJ Delft, The Netherlands}
\affiliation{$^7$Dipartimento di Fisica-Politecnico di Milano, and Instituto di Fotonica e Nanotecnologie-CNR,piazza Da Vinci 32, Milan, Italy}
\date{\today}

\begin{abstract}We demonstrate photon-pair generation in a reverse proton exchanged waveguide fabricated on a periodically poled magnesium doped stoichiometric lithium tantalate substrate. Detected pairs are generated via a cascaded second order nonlinear process where a pump laser at wavelength of 1.55~$\mu$m is first doubled in frequency by second harmonic generation and subsequently downconverted around the same spectral region. Pairs are detected at a rate of 42 per second with a coincidence to accidental ratio of 0.7. This cascaded pair generation process is similar to four-wave-mixing where two pump photons annihilate and create a correlated photon pair.
\end{abstract}

\maketitle


Quantum optics is one of the leading approaches for the implementation of quantum information science protocols \cite{ob-sci-318-1567}.
Recently, the demonstration of quantum-computational gates \cite{po-sci-320-646, po-sci-325-1221} and multimode quantum interference of photons \cite{pe-sci-329-1500} in integrated structures received great interest because it raised the possibility of fabricating future quantum computers on a solid state substrate where photons can be generated, propagate and be manipulated inside optical waveguides. Waveguide based photon-pair sources have also been demonstrated in different materials such as periodically poled lithium niobate \cite{hu-ol-35-1239} (via spontaneous parametric down conversion (SPDC)), in silicon \cite{sh-oe-14-12388} and chalcogenide \cite{xi-apl-98-051101} waveguides (via spontaneous four wave mixing (SFWM)).

The full potential of stability and scalability inherent in the integrated photonic approach can be fully exploited only in a single-chip platform which can incorporate both nonclassical sources and reconfigurable quantum gates together. In this context, ferroelectric crystals like lithium niobate (LN) and lithium tantalate (LT) represent two candidates as host materials for parametric downconversion sources and linear-optics quantum gates. Both LN and LT have a high second order nonlinear coefficient and high quality optical waveguides have been fabricated on both materials through reverse proton exchange \cite{pa-ol-27-179, ma-oe-14-248}. SPDC in a broad wavelength range can be obtained in periodically poled crystals via the quasi-phase-matching technique while fast reconfigurable circuits can be realized by taking advantage of the electro-optic effect of the substrates.
\begin{figure}[!b]
\vspace{-0.30 in}
\centerline{\includegraphics[scale=1]{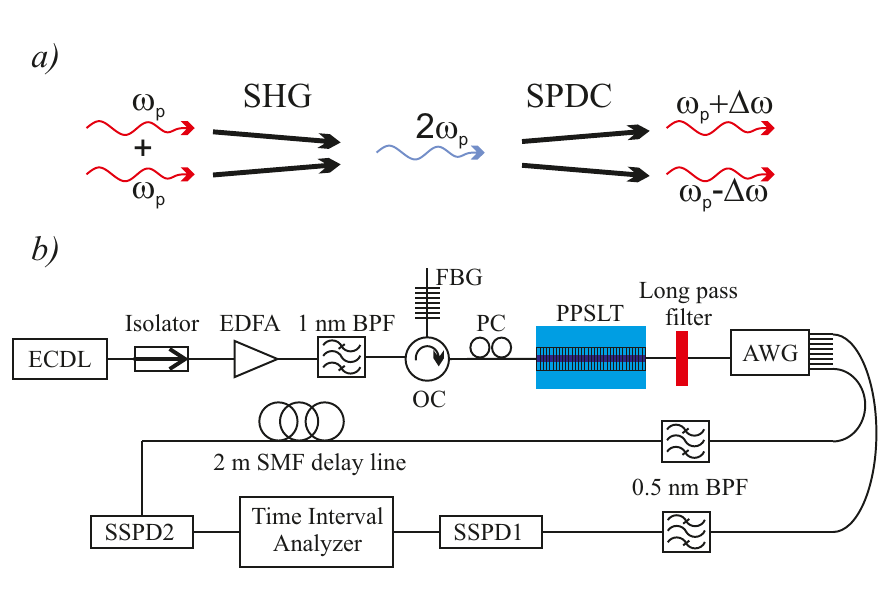}}
\caption{(a)Cascaded second order nonlinear process for the generation of photon pairs on the sidebands of the pump frequency. (b)Scheme of the experimental set-up. ECDL: external-cavity diode laser, EDFA: erbium-doped fiber amplifier. A band-pass filter (BPF) with a full width at half maximum (FWHM) of 1~nm, an optical circulator (OC) and a fiber Bragg grating (FBG, reflection FWHM 0.5~nm) were employed to reduce noise from the ECDL and EDFA. PC: polarization controller, AWG: arrayed waveguide grating, SMF: single-mode fiber, PPSLT: periodically poled stoichiometric lithium tantalate, SSPD: superconducting single photon detector.}
\label{setup}
\end{figure}

Here we demonstrate photon pair generation in a reverse proton exchanged waveguide fabricated on a substrate of periodically poled 1\% MgO doped stoichiometric lithium tantalate (PP:MgSLT) \cite{ma-oe-12-2754}. Although PP:MgSLT has a lower nonlinear coefficient than congruent LN it has a much higher optical damage threshold allowing room temperature operation of the device at pump powers of approximately 200~mW in a single mode waveguide at 1550~nm \cite{lo-ol-31-83}. This simplifies the experimental set-up since a LN waveguide needs to be heated in a crystal oven in order to avoid photorefractive damage. Furthermore PP:MgSLT has a lower coercive field (~1.7~kV/mm) which enables a better quality of the periodic poling and a broader transparency window that extends from 350~nm to $\sim$5~$\mu$m.


\begin{figure}[b]
\centerline{\includegraphics[scale=1]{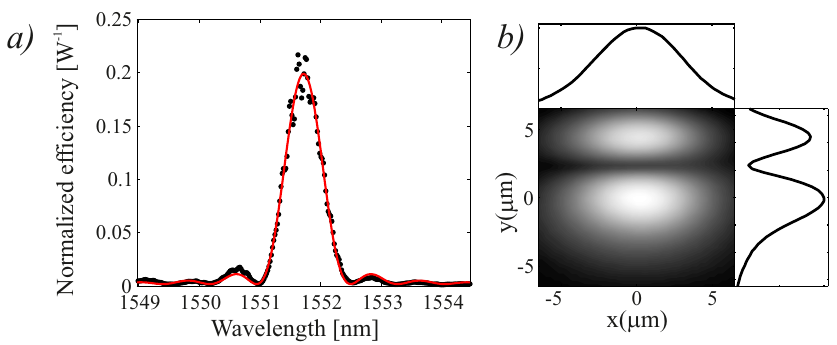}}
\caption{(a)Second harmonic generation as a function of the pump wavelength. The vertical axis shows the ratio between the power at the second harmonic and the square of the power at the fundamental ($P_{2\omega}/P_{\omega}^2$). Solid line is the theoretical fitting and dots are experimental data. (b)Measured intensity profile of the second harmonic mode plotted on a linear arbitrary scale.}
\label{SHG}
\end{figure}

We fabricated a buried waveguide in PP:MgSLT that was single moded at 1550~nm for a channel width of 9~$\mu$m  and had a length of 2.5~cm with insertion losses of 0.7~dB. Correlated photon pairs are generated via a cascaded second order nonlinear process (Fig.~\ref{setup}a): the CW pump mode is phase matched for second harmonic generation (SHG) producing a field at double its frequency that grows quadratically along the waveguide length $z$ according to the equation \cite{pa-ol-27-179}
\begin{equation}
\label{eq.shg}
    P_{2\omega}(z)=\frac{2\pi^2d^2_{eff}}{\lambda_\omega^2\varepsilon_0 c n^2_{2\omega} n_\omega}\frac{z^2}{A_{eff}}P^2_\omega=\eta_{norm}z^2P^2_\omega
\end{equation}
where $d_{eff}=2/\pi d_{33}$ is the effective nonlinear coefficient for quasi-phase-matched SHG, $n_{\omega,2\omega}$ is the refractive index of the material, $\lambda_\omega$ is the pump wavelength, $A_{eff}$ is the effective area of the interaction and  $\eta_{norm}$ is the waveguide efficiency for SHG. Subsequently the photons at $2\omega$ are downconverted into pairs of photons at symmetric sidebands with respect to the pump frequency. The pair production rate in a frequency interval $\Delta\omega\ll2\omega$ and with almost degenerate signal and idler ($\omega_s\approx\omega_i$) can be calculated by adapting the theory described in Ref. \cite{fi-oe-15-7479} for the case where the pump power grows quadratically along the waveguide, leading to
\begin{equation}
\label{eq.spdc}
    C=\frac{d^2_{eff}\eta_{norm}\omega_s^2}{4\pi\varepsilon_0 c^3 n^2_{2\omega} n_\omega A_{eff}}z^4P^2_\omega\Delta\omega.
\end{equation}
The overall process is analogous to SFWM since two pump photons are annihilated for each generated pair at frequencies that are almost degenerate with the pump. 

Figure \ref{setup}b shows the experimental set-up. A tunable external cavity diode laser (ECDL) is amplified in an erbium doped fiber amplifier (EDFA) and wavelength filtered with a 1 nm band-pass filter and a Bragg grating in order to remove amplified spontaneous emission. Vertically polarized light is coupled into the TM$_{00}$ mode of the waveguide which is the only polarization guided in proton exchanged waveguides.

Figure \ref{SHG}a shows the normalized efficiency $P_{2\omega}/P^2_\omega$ for SHG as a function of the pump wavelength. The poling period of the substrate is 21~$\mu$m and corresponds to a phase-matching wavelength of 1551.71~nm when the sample temperature is 27.2$^o$C. In the SHG process the fundamental TM$_{00}$ mode is upconverted into the TM$_{10}$ spatial mode at double the frequency (Fig. \ref{SHG}b). The curve of the SHG power as a function of the pump wavelengths has a FWHM of 0.7~nm which determines an interaction length of 2.2 cm; from data and from the peak value of the curve the effective area between the interacting modes $A_{eff}$~=~141~$\mu$m$^2$ is calculated \cite{pa-ol-27-179}. The conversion efficiency of the $\mathrm{TM_{00}^\omega}\rightarrow\mathrm{TM}_{10}^{2\omega}$ interaction is limited by the spatial overlap between the modes. However the phase matching condition between these two modes is far less critical with respect to inhomogeneities in the refractive index profile of the waveguide than the more efficient $\mathrm{TM_{00}^\omega}\rightarrow\mathrm{TM}_{00}^{2\omega}$ interaction. In our waveguide these inhomogeneities are produced by nonuniform diffusion of hydrogen and lithium during reverse proton exchange and result in a reduced interaction length of the SHG between the $\mathrm{TM_{00}}$ modes to half the sample length cancelling the benefit of a better spatial overlap \cite{bo-tqe-30-2953}.

\begin{figure}[t]
\centerline{\includegraphics[scale=1]{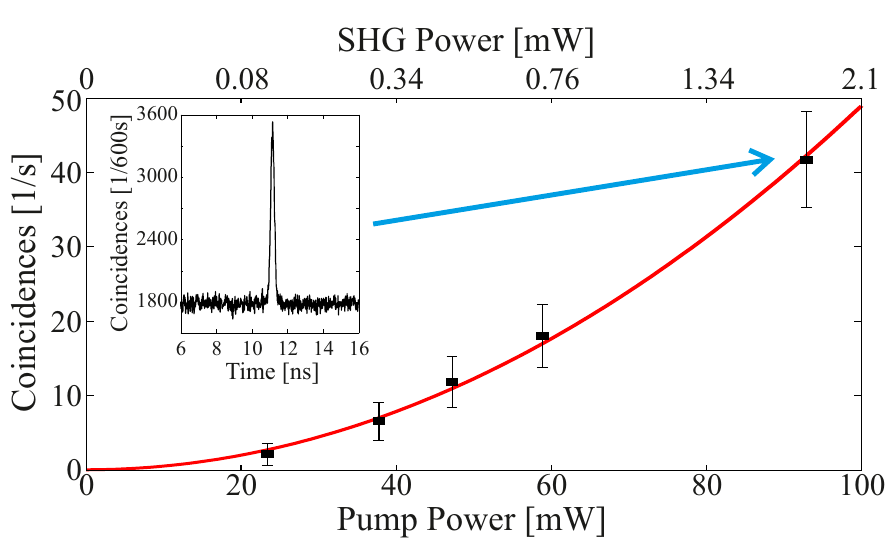}}
\caption{Number of coincidences as a function of the pump power. Dots show the experimental data, the red line is a quadratic fitting. On the top axis coincidences are plotted over the second harmonic power. Inset shows a raw data histogram from the time interval analyzer for a pump power of 93~mW. The peak FWHM is 250~ps and it is the convolution of the time jitter of the measuring set-up and the  time duration of the single photon pulses.}
\label{Pow_Coinc}
\end{figure}

Photon pairs are generated by pumping the waveguide at the phase-matching wavelength for the SHG. The waveguide output is sent into an arrayed waveguide grating (AWG) with 40 channels of 50~GHz FWHM equally spaced by 100~GHz where the different frequencies are coupled into different fibres.
Two symmetric channels with respect to the pump frequency, at 1560.6~nm and 1542.9~nm, are collected as idler and signal photons and passed through a BPF (FWHM=0.5~nm) in order to further reduce the 1551.71~nm pump photon leakage. As the BPF does not provide blocking of 775~nm SHG photons, we put a long-wavelength pass filter before the AWG to block the 775~nm photons.  The photons propagating in the two AWG channels are detected by fiber-coupled superconducting single photon detectors (SSPD) based on NbTiN nanowires \cite{do-apl-93-131101,ta-apl-96-221109}. The detectors have system detection efficiencies of 8\% (SSPD1) and 18\% (SSPD2) at a dark count rate of 1000~s$^{-1}$. One of the channels is optically delayed by a 2~m long single-mode fiber and the output of the detectors is processed by a time interval analyzer (TIA).

\begin{figure}[t]
\centerline{\includegraphics[scale=1]{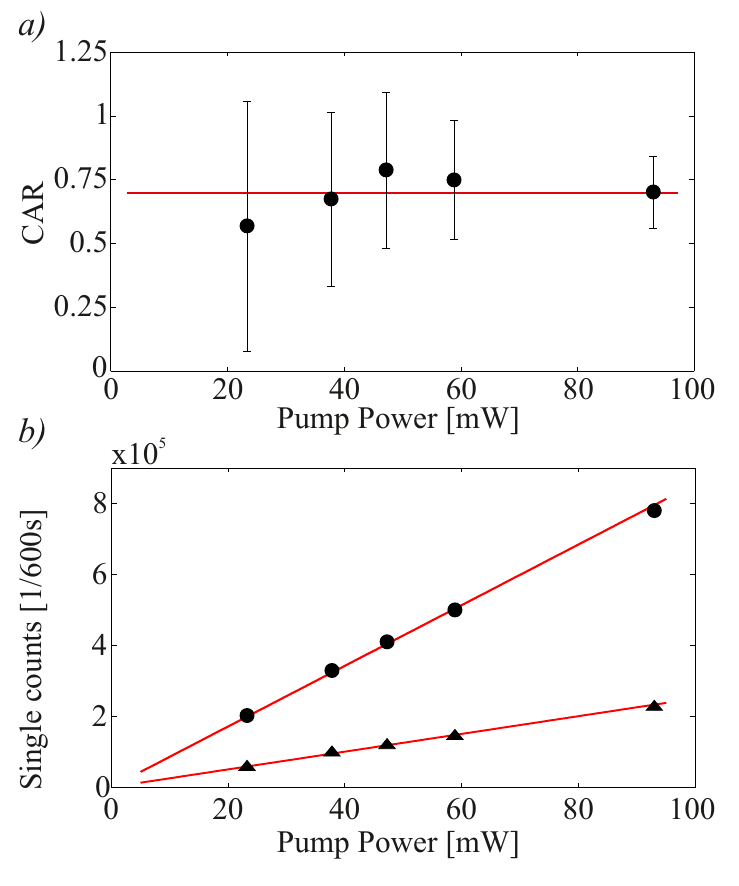}}
\vspace{-0.20 in}
\caption{a) Coincidence to accidentals ratio (CAR) as a function of the pump power with a constant fitting of 0.69. b) Single counts from the two SSPDs for the different pump powers with a linear fit. $(\blacktriangle)$ and $(\bullet)$ are the experimental counts from SSPD1 and SSPD2 respectively.}
\label{CAR}
\end{figure}

Figure \ref{Pow_Coinc} shows the quadratic behaviour of the coincidence rate as a function of the pump power according to Eq.~\ref{eq.spdc}. The inset of Fig.~\ref{Pow_Coinc} shows a trace of the TIA histogram when the waveguide is pumped with 93~mW: the coincidence peak is in correspondence with a 11~ns delay because of the extra 2~m of SMF fiber in the idler channel. Data points of the TIA that are outside the peak are used to measure the number of accidentals per time bin. This number is subtracted from the peak in order to estimate the net coincidences plotted in Fig.~\ref{Pow_Coinc}.
The coincidence to accidentals ratio (CAR, defined as coincidences/accidentals), is shown in Fig. \ref{CAR}a. The CAR value is, within the error bars, independent from the pump power and equal to 0.7.

The source of accidentals is currently under investigation but it most likely depends on a combination of leaked laser light (which include radiation at 980~nm from the EDFA) and Raman generation since the dependence of single counts at the detectors as function of the pump power is linear (Fig.~\ref{CAR}b) and not quadratic as expected from the cascaded nonlinear process. Nevertheless the time resolved measurement implemented with the TIA allows the extrapolation of the net coincidence rate even in the presence of this background.
%
%


In conclusion we have demonstrated correlated photon pairs generation in a PP:MgSLT reverse proton exchanged waveguide using a cascaded second order nonlinear process. The advantages of this device include higher optical damage threshold with respect to LN and lower losses of the waveguide when compared to silicon or chalcogenide devices. We propose that lithium tantalate may be used as a common platform for the implementation of several key components for quantum information technologies using a single fabrication technique. The second order nonlinearity of this material can be used for SPDC photon generation and for electro-optic phase shifters as part of reconfigurable circuits. Higher conversion efficiency can be achieved with a more homogeneous waveguide that uses the interaction $\mathrm{TM_{00}^\omega}\rightarrow\mathrm{TM}_{00}^{2\omega}$ over a longer interaction length in order to generate more second harmonic power and consequently more SPDC pairs.
%
\\
\indent
This work was supported by EPSRC, ERC, QUANTIP, Q-ESSENCE and NSQI, the Australian Research Council Centre of Excellence and Federation Fellowship programs; the Australian Academy of Science's International Science Linkages scheme [GDM]; Royal Society Wolfson Merit Awards [JLO'B];
the Royal Society of London University Research Fellowship [RHH]; RHH acknowledges assistance in constructing the SSPD system from Dr Sae Woo Nam at NIST, USA; the Marie Curie International Incoming Fellowship [ML]; VIDI funding by FOM [SND, TZ and VZ]. CMN, MGT and RHH acknowledge support from EPSRC (UK) and assistance in constructing the SSPDs from Dr S. Nam at NIST, USA.


\bibliography{BibAll}{}
\end{document}